\documentclass[12pt]{article}
\usepackage{sw20lart}

\input tcilatex
\QQQ{Language}{
American English
}

\begin{document}

\title{General Solutions Of The Spherically Symmetric Vacuum Einstein Field
Equations }
\author{Amir H. Abbassi \\
Department of Physics, School of Sciences,\\
Tarbiat Modarres University, P.O.Box 14155-4838,\\
Tehran, I.R.Iran\\
E-mail: ahabbasi@net1cs.modares.ac.ir}
\date{}
\maketitle

\begin{abstract}
According to Birkhoff's theorem the only spherically symmetric solution of
the vacuum Einstein field equations is the Schwarzschild solution. Inspite
of imposing asymptotically flatness and staticness as initial conditions we
obtain that these equations have general solutions with the Schwarzschild
metric as merely a special and simplest form of them. It is possible to have
perfect and smooth metrics with the same Newtonian and post-Newtonian limits
of Schwarzschild by a convenient and correct selection.

\end{abstract}

\newpage\ Spherical symmetry requires the dependence of line element $ds^2$%
only on the following rotatinal invariants [1-2],

\[
t\;,\;dt\;,\overrightarrow{x}\cdot d\overrightarrow{x}^2=rdr\;,\;d 
\overrightarrow{x}^2=dr^2+r^2(d\theta ^2+\sin {}^2\theta \;d\varphi ^2) 
\]
So it can be written

$\;$%
\begin{equation}  \label{1}
ds^2=B(r,t)dt^2-A(r,t)dr^2-2C(r,t)dtdr-D(r,t)(d\theta ^2+\sin {}^2\theta
\;d\varphi ^2)
\end{equation}

There is a common belief that the line element can be transformed to the
following standard form by a suitable coordinate transformation. Some texts
start from here e.g.\cite{3}.

$\;$%
\begin{equation}
ds^2=B^{^{\prime }}(r^{^{\prime }},t^{^{\prime }})dt^{^{\prime
}2}-A^{^{\prime }}(r^{^{\prime }},t^{^{\prime }})dr^{^{\prime
}2}-r^{^{\prime }2}(d\theta ^{^{^{\prime }}2}+\sin {}^2\theta ^{^{\prime
}}\;d\varphi ^{^{\prime }2})
\end{equation}

Taking advantage of the mentioned form to compute the field equations it can
be shown that it is necessarily static and has a unique Schwarzschild
solution[4-5], as required by Birkhoff's theorem [6-7]. This means that the
other solutions are just different forms of this metric which are related by
coordinate transformation. What indeed is flawing the reasoning, is that the
change of the coordinate $r\rightarrow \sqrt{D(r,t)}$ with new parameter
having the same range of $r,$ is not generally true. Because as it turns out
as in the Schwrazschild case, the components of the metric may even change
their sign in different parts of space-time. So when there is no information
about the functional form of $D$ there is no guarantee that$\sqrt{D}$ to be
real in the whole range of $r$ and also, its range is exactly the same as $r$
which is between zero and infinity. Accordingly we believe that the steps
which follow to arrive at the standard form are not justified. Since it is a
hard task to solve the vacuum field equations with the general form of the
metric (1) we restrict our investigation to asymptotically flat and static
space-time by convention i.e.

\begin{equation}
ds^2=B(r)dt^2-A(r)dr^2-D(r)(d\theta ^2+\sin {}^2\theta \;d\varphi ^2)
\end{equation}
This metric tensor has the nonvanishing components

\begin{equation}
g_{tt}=-B(r)\;\;\;\;,\;\;\;\;g_{rr}=A(r)\;\;\;,\;\;\;g_{\theta \theta
}=D(r)\;\;\;,\;\;\;g_{\varphi \varphi }=D(r)\sin {}^2\theta
\end{equation}
with functions $A(r)\;,\;B(r)\;$and $D(r)$ that are to be determined by
solving the field equations. The nonvanishing contravariant components of
the metric are :

\begin{equation}
g^{tt}=-B^{-1}\;\;,\;\;\;g^{rr}=A^{-1}\;\;,\;\;g^{\theta \theta
}=D^{-1}\;\;,\;\;g^{\varphi \varphi }=D^{-1}\sin {}^{-2}\theta
\end{equation}

The metric connection can be computed by the use of (4) and (5) from the
usual definition. Its only nonvanishing components are:

\begin{equation}
\begin{array}{llll}
\Gamma _{rr}^r=\frac{A^{^{\prime }}}{2A}\; & \Gamma _{\theta \theta }^r=- 
\frac{D^{^{\prime }}}{2A} & \Gamma _{\varphi \varphi }^r=-\sin {}^2\theta \; 
\frac{D^{^{\prime }}}{2A} & \Gamma _{tt}^r= \frac{B^{^{\prime }}}{2A} \\ 
\Gamma _{r\theta }^\theta =\frac{D^{^{\prime }} }{2D} & \Gamma _{\varphi
\varphi }^\theta =-\sin \theta \;\cos \theta &  &  \\ 
\Gamma _{r\varphi }^\varphi =\frac{D^{^{\prime }}}{2D} & \Gamma _{\varphi
\theta }^\varphi =\cot \theta &  &  \\ 
\Gamma _{tr}^t=\frac{B^{^{\prime }}}{2B} &  &  & 
\end{array}
\end{equation}
where primes stand for differentiation with respect to $r$. With these
connections the Ricci tensor can be obtained.

\begin{equation}
R_{rr}=\frac{B^{^{\prime \prime }}}{2B}-\frac{B^{^{\prime }}}{4B}(\frac{%
A^{^{\prime }}}A+\frac{B^{^{\prime }}}B)-\frac{A^{^{\prime }}D^{^{\prime }}}{%
2AD}+\frac{D^{^{\prime \prime }}}D-\frac{D^{^{\prime }2}}{2D^2}
\end{equation}

\begin{equation}
R_{\theta \theta }=-1+\frac{D^{^{\prime }}}{4A}(-\frac{A^{^{\prime }}}A+ 
\frac{B^{^{\prime }}}B)+\frac{D^{^{\prime \prime }}}{2A}
\end{equation}

\[
R_{\varphi \varphi }=\sin {}^2\theta \;R_{\theta \theta } 
\]

\begin{equation}
R_{tt}=-\frac{B^{^{\prime \prime }}}{2A}+\frac{B^{^{\prime }}}{2A}(\frac{%
A^{^{\prime }}}A+\frac{B^{^{\prime }}}B)-\frac{B^{^{\prime }}D^{^{\prime }}}{%
2AD}
\end{equation}

\[
R_{\mu \nu }=0\;\;\;\;for\;\mu \neq \nu 
\]
The Einstein field equations for vacuum are $R_{\mu \nu }=0$. Dividing $%
R_{rr}$ by $A$ and $R_{tt}$ by $B$ and putting them together we get

\begin{equation}
-\frac{D^{^{\prime }}}{2AD}(\frac{A^{^{\prime }}}A+\frac{B^{^{\prime }}}%
B)+\frac 1A(\frac{D^{^{\prime \prime }}}D-\frac{D^{^{\prime }2}}{2D^2)}=0
\end{equation}
Multiplying (10) by $\frac{2AD}{D^{^{\prime }}}$ gives

\begin{equation}
\frac{A^{^{\prime }}}A+\frac{B^{^{\prime }}}B=\frac{2D^{^{\prime \prime }}}{%
D^{^{\prime }}}-\frac{D^{^{\prime }}}D
\end{equation}
Now let integrate (11) with respect to $r$ and find

\begin{equation}
AB=C_1\frac{D^{^{\prime }2}}D
\end{equation}
where $C_1$is a constant of integration which can be fixed by requiring that 
$D$ asymptotically approaches to $r^2$ and $A$ and $B$ to one. This will fix 
$C_1$ to $\frac 14$ by (12), thus

\begin{equation}
AB=\frac{D^{^{\prime }2}}{4D}
\end{equation}
Now using (9) and dividing the field equation $R_{tt}=0$ by $\frac{%
B^{^{\prime }}}{2A}$ we obtain

\begin{equation}
-\frac{B^{^{\prime \prime }}}{B^{^{\prime }}}+\frac 12(\frac{A^{^{\prime }}}%
A+\frac{B^{^{\prime }}}B)-\frac{D^{^{\prime }}}D=0
\end{equation}
Substituting (11) in (14) we get

\begin{equation}
\frac{B^{^{\prime \prime }}}{B^{^{\prime }}}+\frac{3D^{^{\prime }}}{2D}= 
\frac{D^{^{\prime \prime }}}{D^{^{\prime }}}
\end{equation}
The next step is to integrate (15) with respect to $r$ which gives

\begin{equation}
B^{^{\prime }}D^{\frac 32}=C_2D^{^{\prime }}
\end{equation}
where $C_2\;$is a constant of integration. Dividing (16) by $D^{\frac 32}$
and taking another integration with respect to $r$ we get

\begin{equation}
B=-2C_2D^{-\frac 12}+C_3
\end{equation}
where C$_3$ is another constant of integration. $C_2$ and $C_3$ can be fixed
by considering the Newtonian limit of $B$ which gives $C_3=1$ and $C_2=M$ ($%
G=c=1$). Thus

\begin{equation}
B=1-2MD^{-\frac 12}
\end{equation}
$A$ may be computed by (12) and (18). It is

\begin{equation}
A=\frac{\frac{D^{^{\prime }2}}{4D}}{1-2MD^{-\frac 12}}
\end{equation}
It is a natural expectation that the functional form of $D$ to be fixed by
using the $\theta \theta $ component of the field equation, that is $%
R_{\theta \theta }=0.$ But using (18) , (19) and (11) it turns out that the
equation $R_{\theta \theta }=0$ will become an identical relation of zero
equal to zero for any arbitrary function of $D$ which is only restricted to
the following constraints

\begin{equation}
D\rightarrow r^2\;\;\;\;\;\;if\;\;\;\;\;\;\;r\rightarrow \infty \;\cup
\;M\rightarrow 0
\end{equation}
This means that $D$ has the functional form

\begin{equation}
D(r,M)=r^2\;f(\frac Mr)
\end{equation}
where $f(0)=1$. There is not any obligation that $D(0,M)$ becomes zero.
Since the behavior of $f$ approaching infinity i.e. r$\rightarrow 0$ is
ambiguous then the transformation $r\rightarrow r^{^{\prime }}=\sqrt{D(r,M)}%
=r\sqrt{f(\frac Mr)}$ does not specify the range of $r^{^{\prime }}$ at all.
It is clear that the Schwarzschild solution is a special case of the general
solution by choosing $\;f=1,$ indeed the simplest case. This does not based
on any physical fact. This unjustified choice which is not even stand on any
fundamental reasoning, is the source of the trouble of singularity in some
parts of space-time and of course can be avoided. The appearance of square
root of $D$ in the final solution reveals that $D$ cannot adapt negative
values otherwise the metric becomes complex which physically is not
acceptable. Then we may perform the following coordinate transformation

\begin{equation}
r\;\text{with the range }(0,+\infty )\;\rightarrow \;r^{^{\prime
}}=D(r,M)^{\frac 12}\;\text{with the range }(D(0,M)^{\frac 12},0)
\end{equation}
where $D(0,M)$ may be any arbitrarily real non-negative number. Since $D$
has $\left[ L^2\right] $ dimension and $D(0,0)$ is equal to zero and also $M$
is the only natural parameter of the system which has length dimension we
may conclude that the general form of $D(0,M)$ is

\begin{equation}
D(0,M)^{\frac 12}\;=\;\alpha (M)M
\end{equation}
where $\alpha $ is a dimensionless parameter which for simplicity may take
it as a constant independent of $M$. Now we make another coordinate
transformation

\begin{equation}
r^{^{\prime }}\;\rightarrow \;r^{^{\prime \prime }}=r^{^{\prime }}-\alpha M
\end{equation}
The range of new radial coordinate is from zero to infinity. Dropping primes
the final form of the metric becomes

\begin{equation}
ds^2=(1-\frac{2M}{r+\alpha M})dt^2-\frac{dr^2}{1-\frac{2M}{r+\alpha M}}%
-(r+\alpha M)^2(d\theta ^2+\sin ^2d\varphi ^2)
\end{equation}
Thus $\alpha $ is an arbitrary constant which its different values define
the members of our general solutions family. The only condition on $\alpha $
is that it should not be much much bigger than one, otherwise it would
contradict with Newtonian mechanics predictions. The corresponding metrics
related to the values of $\alpha $ bigger than 2 are all regular in the
whole space. The actual metric of the spacetime of course is merely a
special member of this class but not necessarily the Schwarzschild metric($%
\alpha =0$). Specifying this requires further information about the actual
properties of spacetime at Schwarzschild scales which at this time there is
no access to such data.

\begin{center}
\textbf{Test for Equivalence}
\end{center}

Someone may be taken in by the apparent form of the general solution that
these are just the usual Schwarzschild solution with $\;r$ \ replaced by $%
\;r+\alpha M$ and conclude that they are not new. In the I clarify this
point and show that it is not so.

\noindent 1- In Schwarzschild solution the range of radial coordinate $r$ is
between $0\;$to$\;\infty $ and the center of symmetry is at $r=0$. Replacing 
$r$ by $\acute{r}+\alpha M$ does not lead to the new solution because the
range of the transformed radial coordinate is between $-\alpha M$ \ to $%
\infty .$ Of course negative values for conventional radial coordinate is
meaningless and the center of symmetry is located at $\acute{r}=-\alpha M$.
This is not identical to the new solution because the range of radial
coordinate is between $0$ to $\infty $ and the center of symmetry is at $%
\acute{r}=0$. On the other hand in the new metric replacing $r+\alpha M= 
\acute{r}$ does not lead to Schwarzschild solution because the range of $%
\;r\;$here is between $0$ to $\infty $ and the center of symmetry is at $r=0$%
. While the range of the transformed radial coordinate is between $\alpha M$
to $\infty $ and the center of symmetry is at $\acute{r}=\alpha M$. This
evidently is different from Schwarzschild solution. Though the extension of $%
\acute{r}$ to values smaller than $\alpha M$ is physically meaningless
because a point by definition has no internal structure to be extended
inside of it, let us consider such a mathematical hypothetical spacetimes.
Even these are essentially different from Schwarzschild spacetime because in
Schwarzschild the point mass $M$ is at $r=0$ while in these spacetimes it is
located at $r=\alpha M$.

\noindent 2- The center of spherical symmetry that is the position of point
mass $M$ is a common point between the Schwarzschild spacetime and the
presented spherically symmetric vacuum spacetimes in this manuscript. as it
has been shown the field equations and the given boundary conditions are not
sufficient to fix $\alpha $. Thus if we take $\alpha \neq 0$ , then these
solutions will not be singular at the center of symmetry while the
Schwarzschild spacetime possesses an intrinsic singularity at the center of
symmetry. If these new metrics were isometric to Schwarzschild metric they
should be singular too, because coordinate transformation cannot change the
intrinsic properties of spacetime. This clearly shows that the presented
metrics are not Kottler solutions of Schwarzschild spacetime.

\noindent 3- The presented general solution and the Schwarzschild solution
have exactly the same space extension. Making use of Cartesian coordinate
system as frame of reference will elucidate this fact. It turns out that all
components have the same range $(-\infty ,+\infty ).$ 

\noindent 4- Let us consider hypothetically spacetimes which possesses
different lower bound for the surface area of a sphere. Obviously they have
different geometrical structures and present different physics.

\noindent 5- The zone of $r$ of the order of Schwarzschild radius is the
domain in which gravitational field is tremendously strong and
conventionally we have to give up our common sense and replace the character
of $r$ and $t$. So it is not surprising to have a geometry completely
different.

\bigskip\ 

We may conclude the discussion with this statement that in contrast with
Birkhoff's theorem the vacuum Einstein field equation spherical solution is
not either automatically static nor is uniquely Schwarzschild.

\newpage\ \


\begin{thebibliography}{9}
\bibitem{1}  Weinberg,S. \emph{Gravitation and Cosmology }, John
Wiley,p335(1972).

\bibitem{2}  D'Inverno,R. \emph{Introducing Einstein's Relativity},
Clarendon Press,p185 (1992).

\bibitem{3}  Rindler,W.\emph{Essential Relativity },Springer-Verlag,
p136(1977).

\bibitem{4}  Joshi,P.S.\emph{GlobalAspects in Gravitation\&Cosmology}%
,ClarendonPress,p66 (1993).

\bibitem{5}  D'Eath,P.D.\emph{Black Holes,Gravitational Interactions}%
,Clarendon Press,p19 (1996).

\bibitem{6}  Birkhoff,G.D. \emph{Relativity and Modern Physics}, Harvard
Univ.Press(1923).

\bibitem{7}  Heusler,M.\emph{BlackHoleUniquenessTheorems},Camb.Univ.Press,
p14(1996).
\end{thebibliography}
\end{document}